\documentstyle[latexsym,epsf,amsfonts,prd,aps,twocolumn]{revtex}
       
\newcommand{\gl}[1]{(\ref{#1})}
\newcommand{\sla}[1]{\not\! #1 \,}
\newcommand{\be}[1]{\begin{equation}\label{#1}}
\newcommand{\ee}{\end{equation}}
\newcommand{\ba}[1]{\begin{eqnarray}\label{#1}}
\newcommand{\ea}{\end{eqnarray}}
 
\newcommand{\ket}[1]{\left|#1\right>} 
\newcommand{\bra}[1]{\left<#1\right|} 
\newcommand{\nn}{\nonumber\\} 
 
\newcommand{\braket}[2]{\left<#1|#2\right>} 

\newcommand{\eps}{\varepsilon}

\begin{document} 
 
\title{On the anomalous dimension for the transversity distribution}
\author{Michael Meyer-Hermann$^{1,*}$,
Ralf Kuhn$^{1,2}$, and Ralf Sch\"utzhold$^1$}
\address{$^1$Institut f\"ur Theoretische Physik, Technische  Universit\"at
Dresden, D-01062  Dresden, Germany\\
$^2$Max-Planck-Institut f\"ur Physik komplexer Systeme, 
D-01187 Dresden, Germany\\
$^*$Electronic address: {\tt meyer-hermann@physik.tu-dresden.de}}
\date{\today}
\maketitle

\begin{abstract} 
We show that a standard calculation of the
splitting function for the
nonsinglet structure function $h_1$ does not lead 
to the expected result \cite{Art90}. 
The calculation is compared to the corresponding
derivation of the splitting function for the nonsinglet
polarized structure function $g_1$.
We analyze possible explanations for the unexpected
result and discuss its implications.
\end{abstract} 
   

\section{Introduction}%%%%%%%%%%%%%%%%%%%%%%%%%%%%%%%%%%%%%%%%%%%%%%%%%%%%%

In the Drell-Yan process a quark-antiquark pair is annihilated
to a virtual photon, which produces lepton pairs.
The corresponding cross section contains a contribution
from quarks of different chirality: a quark originating
from one nucleon may return to this nucleon in the squared
scattering amplitude with its chirality flipped. This is
possible because the intermediate photon does not remember
the distribution of quark chirality on the scattering nucleon.
This gives rise to the leading twist
chiral odd distribution function $h_1$, which
first appeared in the discussion of the transverse polarized
Drell-Yan process \cite{Ral79,Jaf91,Cor92,Tan95}.
One should emphasize that such chiral odd quark distribution functions
do not appear in totally inclusive deep inelastic scattering
because the quark chirality is conserved at each vertex
of the scattering process. 

The quark helicity distribution
$g_1$ has a partonic interpretation in the chiral basis.
This is not the case for $h_1$ which gets a partonic
interpretation only by switching to the transversity
basis \cite{Gol76}. Looking on transversely polarized
nucleons $h_1$ is interpreted as the probabilistic asymmetry
of finding quarks in eigenstates of the transverse Pauli-Lubanski vector
with eigenvalue $\pm 1/2$. One should be aware, that
$g_1$ looses its partonic interpretation in the transversity 
basis. 

There are several attempts to measure the transversity
distribution $h_1$, see e.g.~\cite{Cor92,Col94,Jaf95,Kor00,Ma00}. 
For general reasons the transversity distribution has to obey
restrictions as Soffers inequality \cite{Sof95}
for twist-2 distribution functions 
$2|h_1(x)|\leq f_1(x)+g_1(x)$.

The anomalous dimension of $h_1$ and the corresponding 
splitting function can be found in literature 
\cite{Art90,Shi81}. This result is found by direct
evaluation of the splitting function. Nevertheless the
behavior at Bjorken-$x$ equal to 1 remains unfixed in this
procedure. In the case of the helicity distribution $g_1$
the behavior for $x=1$ is fixed by a corresponding sum
rule. For $h_1$ such a sum rule is lacking.
So the full $h_1$ splitting function was determined by
a general argument, which demands the splitting function 
of $h_1$ to
behave exactly as the splitting function of $g_1$ for $x\uparrow 1$.
To our knowledge, no explicit calculation (without additional 
arguments) of the full splitting
function of the transversity distribution exists until now.

This paper is organized as follows.
First, the forward scattering amplitudes corresponding
to the helicity and transversity operator are defined.
We derive the leading order anomalous dimension of $g_1$ and 
$h_1$ using a dispersion relation
and show that the behavior of the 
splitting function for $h_1$  
does not match the expected result 
found in literature in the limit $x\uparrow 1$.
We discuss possible explanations and implications of our
result.

\section{Definitions}

\subsection{Helicity operator}

The forward scattering amplitude 
\ba{g1Tmu}
T_{\mu\nu}(p,q,s)
&=&
\frac{i}{2}
\int d^4y\,e^{iqy}
\times
\nonumber\\
&&
\bra{p,s} {\cal T}\left[j_{\mu 5}(y) j_\nu(0)
 + j_\nu(y) j_{\mu 5}(0) \right]
\ket{p,s}
\ea
contributes to the nonsinglet
polarized structure function 
$g_1(x,Q^2) := g_1^p(x,Q^2) - g_1^n(x,Q^2)$, where
$j_{\nu}(y) = \overline{\psi}(y) \gamma_\mu \psi(y)$
is a vector current and
$j_{\mu 5}(y) = \overline{\psi}(y) \gamma_\mu \gamma_5 \psi(y)$
denotes an axial-vector current.
$g_1(x,Q^2)$ depends on all possible Lorentz invariants $Q^2=-q^2$ and
Bjorken $x$ with $x=Q^2/2p\cdot q$. Here $p$ and $q$ are the
four-momenta of the hadron and virtual photon, respectively.
The latter defines the momentum scale of the scattering process.
${\cal T}$ denotes the time-ordered product
and summation over flavor indices is assumed.
In order to get the part relevant for $g_1$ one has to project
the forward scattering amplitude on the correct
Lorentz structure:
\be{Teg1}
g_1
\frac{i\varepsilon_{\mu\nu\lambda\rho} q^\lambda s^\rho}
{p\cdot q} 
\;=\;
\left.
\frac{1}{\pi}\,
{\rm Abs}\, 
T_{\mu\nu} 
\right|_{\mbox{$\displaystyle
\frac{i\varepsilon_{\mu\nu\lambda\rho} q^\lambda s^\rho}
{p\cdot q}$}}
\,.
\ee
Abs denotes the absorptive part, i.e. the discontinuity accross a cut
\cite{Mut87}.

If higher twist operators are neglected,
the moments of the polarized structure function
can be identified
with the twist-2 operator matrix elements $A_{g_1,n}$
in the framework of the operator product expansion:
\ba{Mg1}
M_{g_1,n}
 &=& \int\limits_0^1 dx\;x^n g_1(x) \nonumber\\
 &=& C_{g_1,n} A_{g_1,n}
\,.
\ea
Note, that we dropped the $Q^2$- and the renormalization scale
dependence, which are not important for our purpose. 
$C_n$ is the Wilson-coefficient which depends on the
renormalization scale and the strong coupling. 
It contains the full perturbation series of leading twist.
For example the coefficient for $n=0$ corresponds
to the Bjorken sum rule.

\subsection{Transversity operator}

Let us define the following forward scattering amplitude
(see \cite{Iof95,Kir97,Mey99}):
\ba{h1Tmu}
T_\mu(p,q,s)
&=&
\frac{i}{2}
\int d^4y\,e^{iqy} \times
\nonumber\\
&&
\bra{p,s} {\cal T}\left[j_{\mu 5}(y) j_S(0)
 + j_S(y) j_{\mu 5}(0) \right]
\ket{p,s}
\,.
\ea
$j_{\mu 5}(y)$ is a nonsinglet axial-vector current as for $g_1$ and
$j_S(y) = \overline{\psi}(y) \psi(y)$ a nonsinglet
scalar current.
Again, summation over flavor indices is assumed.
This operator defines a structure function, which
describes vertices without conservation
of chirality. Such vertices do not
exist in QCD. Nevertheless, quark chirality may
be flipped in Drell-Yan processes, in which
quarks of different chirality may be
connected to one virtual photon. This process is not
suppressed by powers of $Q^2$ for large momentum transfers
such that this contribution to the cross section
is of leading twist. Consequently, the structure
function $h_1$ defined by Eq.~\gl{h1Tmu} may be connected
to the tranversity distribution.
Indeed, it was shown on leading twist level that
$h_1$ can be identified with the
transversity distribution \cite{Iof95}
(cf.~the Appendix).

As for the polarized structure function, we
get the contributions of the forward scattering
amplitude to the
structure function $h_1$ by projection 
on the corresponding Lorentz structure:
\be{Teqh1}
h_1(x) s_{\bot\mu} 
\;=\;
\left.
\frac{1}{\pi}\,
{\rm Abs}\, T_\mu
\right|_{\mbox{$\displaystyle s_{\bot\mu}$}}
\,.
\ee
The corresponding moments are defined as in \cite{Jaf91}
\ba{Mh1}
M_{\perp,n}
&=&
\int\limits_0^1 dx\,x^n
\left[h_1(x) + (-)^n h_1(-x)\right]\nn
&=&
C_{\perp,n} A_{\perp,n}
\,.
\ea
The last term again corresponds to the leading twist
contribution in the operator product expansion with
the Wilson coefficient $C_{\perp,n}$ and the chiral odd
operator matrix element $A_{\perp,n}$.

\section{Derivation of splitting functions}
\label{SplitCalc}
\subsection{Wick expansion}

\begin{center}
\begin{figure}
\label{figures}
\begin{minipage}[t]{4.2cm}
\epsfxsize=4cm\epsffile{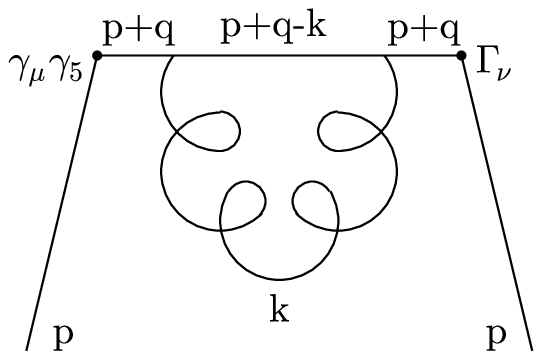}
\end{minipage}
\begin{minipage}[t]{4.2cm}
\epsfxsize=4cm\epsffile{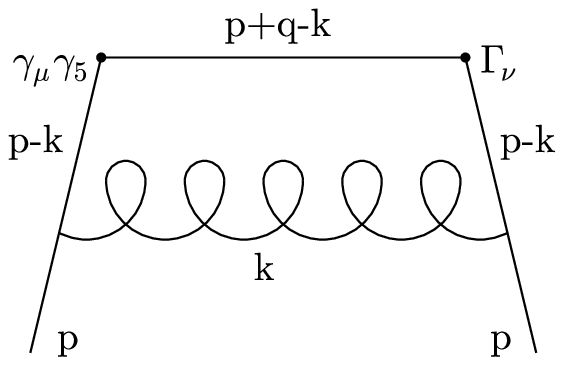}
\end{minipage}
\begin{minipage}[t]{4.2cm}
\epsfxsize=4cm\epsffile{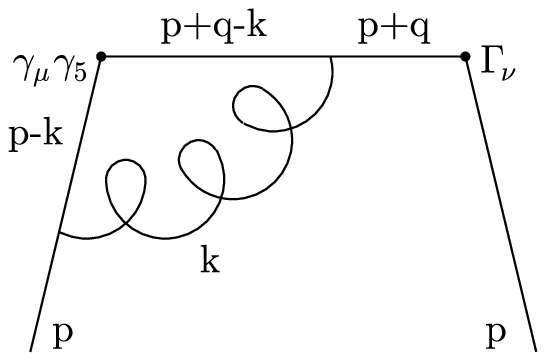}
\end{minipage}
\begin{minipage}[t]{4.2cm}
\epsfxsize=4cm\epsffile{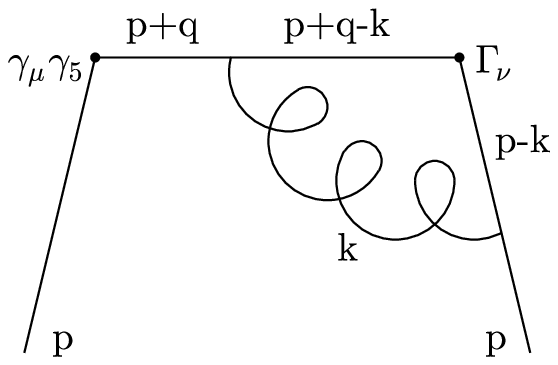}
\end{minipage}
\caption{The Feynman graphs contributing to the forward
scattering amplitude with $\Gamma_\nu=1,\gamma_\nu$,
respectively: The upper graphs represent the selfenergy (left)
and the boxgraph (right) and the lower ones depict the vertex
corrections.}
\end{figure}
\end{center}

The splitting functions for the operators in Eqs.~\gl{g1Tmu}
and \gl{h1Tmu} are determined by the divergent part of
the forward scattering amplitudes. 
By Wick expansion of the current products we find 32 Feynman
graphs contributing to the splitting functions.
These diagrams can be divided into two classes which transform
into each other by substituting the gluon momentum $k\rightarrow -k$.
Since the gluon loop integrals are symmetric it is sufficient to
consider 16 graphs.
The forward scattering amplitude $T$ is real on the real axis, i.e.
$T^*=T$. The remaining 16 graphs split up into 8 diagrams and their
complex conjugates -- which contain the same loop integral and a
Lorentz structure in inverse order. 
The 8 relevant diagrams left consist of 4 diagrams (given below)
with different
loop integrals and the corresponding exchange graphs. The latter
are obtained by substituting the momentum of the virtual photon
$q\rightarrow -q$ and interchanging the external vertices
simultaneously.
The Feynman diagrams are depicted in 
Fig.~\ref{figures}.
The amplitude associated to the selfenergy reads
\ba{self}
{\cal M}_S&=&-C_F\,\overline{u}(p) 
\Gamma_\nu 
%\frac{\sla{p}+\sla{q}}{(p+q)^2}
\frac1{\sla{p}+\sla{q}+i\eta}
\gamma_\rho \gamma_\beta \gamma^\rho 
\times
\nonumber\\
&&
\frac1{\sla{p}+\sla{q}+i\eta}
\gamma_\mu \gamma_5
u(p)\,I_S^\beta
\,,
\ea
where $I_S^\beta$ is given in the Appendix.
In analogy we get for the vertex corrections
\ba{vertexr}
{\cal M}_{V1} &=& C_F\,\overline{u}(p)
\gamma_\rho \gamma_\alpha \Gamma_\nu \gamma_\beta \gamma^\rho
\frac1{\sla{p}+\sla{q}+i\eta}
\times
\nonumber\\
&&
\gamma_\mu \gamma_5
u(p)\,I_V^{\alpha\beta}
\ea
and
\ba{vertexl}
{\cal M}_{V2} &=&-C_F\,\overline{u}(p)\Gamma_\nu
\frac1{\sla{p}+\sla{q}+i\eta}
\times
\nonumber\\
&&
\gamma_\rho \gamma_\beta
\gamma_\mu \gamma_5
\gamma_\alpha \gamma^\rho
u(p)\,I_V^{\alpha\beta}
\,,
\ea
and finally for the boxgraph
\ba{box}
{\cal M}_{B} &=&
C_F\,
\overline{u}(p) 
\gamma_\rho
\gamma_\alpha \Gamma_\nu \gamma_\beta \gamma_\mu \gamma_5
\gamma_\lambda \gamma^\rho
u(p)\,I_B^{\alpha\beta\lambda}
\,.
\ea
The integrals $I_V^{\alpha\beta}$ and $I_B^{\alpha\beta\lambda}$
can be found in the Appendix too.
The exchange graphs are not shown.
$\Gamma_\nu$ stands for
$\gamma_\nu$ or for identity accounting for
the helicity or transversity operator, respectively.
We have to isolate the
divergent parts contributing to the Lorentz structures which
correspond to $g_1$ or $h_1$.

\subsection{Dispersion relation}
\label{secdisp}

In order to calculate the contributions of the above Feynman diagrams
to the splitting functions we have to consider terms like
\be{Lori}
\frac1{\sla{p}+\sla{q}+i\eta}=
\frac{\sla{p}+\sla{q}-i\eta}{Q^2(\omega-1)+\eta^2}
\ee
with
\be{omega}
(p+q)^2=-Q^2(1-\omega) 
\quad {\rm and} \quad
\omega=1/x
\,.
\ee
The limit $\eta\downarrow 0$ cannot be taken easily at this
stage of calculation. For example the Dirac identity
\be{eta}
\lim_{\eta\downarrow 0}\frac1{\chi\pm i\eta} = 
{\cal P}\left(\frac1\chi\right)\mp i\pi\delta(\chi)
\ee
is not sufficient to deduce the resulting distributions.
Although the imaginary part of the r.h.s. of Eq.~\gl{Lori} generates
some $\delta$-type distribution
\ba{impart}
\lim_{\eta\downarrow 0}\frac{-i\eta}{Q^2(\omega-1)+\eta^2}&=&
- i\pi\delta\left(\sqrt{Q^2(\omega-1)}\right)
\nonumber \\
&=&
- 2i\pi \sqrt{Q^2(\omega-1)}\,\delta(Q^2(\omega-1))
\nn
\ea
this distribution does not contribute to the moments
in Eqs.~\gl{Mg1} and \gl{Mh1} owing to the measure $dx$.

We deduce the singularity structure of the splitting functions
using the well-known dispersion relation (see e.g.~\cite{Mut87})
\ba{disp}
T(x)
&=&
\sum\limits_{n=0}^\infty \left(\frac1{x}\right)^{n+1} 
\int\limits_{-1}^1 dy\,
y^n \,\frac1{\pi} 
{\rm Abs}\,T(y) \nn
&=&
\sum\limits_{n=0}^\infty \left(\frac1{x}\right)^{n+1} \int\limits_0^1 dy\,
y^n \times\nn
&&\frac1{\pi} 
\left({\rm Abs}\,T(y) + (-1)^n {\rm Abs}\,T(-y) \right)
\,.
\ea
$T$ denotes the forward scattering amplitude accounting for
$g_1$ and $h_1$, i.e. $T_{\mu\nu}$ and $T_\mu$,
respectively.
These amplitudes are analytic functions of $Q^2$ and $x$.
They possess a pole at $x=1$ and a cut on the real axis
for $x<1$. Through the absorptive part ${\rm Abs}\,T$ we isolate
the discontinuity accross the cut. The dispersion relation
Eq.~\gl{disp} connects the behavior of $T$ in the regular (but
unphysical) region $x>1$ with the physical region $x<1$, i.e.
the coefficients of the Laurent expansion for large $x$ correspond
to the moments of the structure functions.
Having expanded the amplitude $T$, the full information about
the singularity structure is contained in the corresponding 
coefficients. In this representation the limit $\eta\downarrow 0$
can be taken for each coefficient separately. 
In this way the anomalous dimensions are calculated.
By taking the
inverse Mellin transformation we find the 
corresponding splitting functions.

In the case of $g_1$ crossing symmetry holds, so that 
Eq.~\gl{disp} gives rise to even moments only. Odd moments
cancel identically and there is no way of getting information
on the missing moments. Nevertheless, in order to calculate
the splitting function we will have to evaluate the inverse
Mellin transformation of the anomalous dimension, which,
for this reason should be known for all moments. 
As it will become evident later, our main interest lies on the
singularity structure for $x\uparrow1$, in particular we are interested
in contributions like $\delta(1-x)$ to the forward scattering
amplitude $T$.
The $\delta$-distribution is element of the Schwartz-Sobolev
space ${\mathfrak S}_{-1}\left([0,1]\right)$, cf.~\cite{Sch66}.
Its domain of definition is the dual 
space ${\mathfrak S}_{1}\left([0,1]\right)$,
i.e. $\braket{\delta_\chi}{f}=f(\chi)$ exists for all functions $f$
that are element of 
${\mathfrak S}_{1}\left([0,1]\right) \subset L_2\left([0,1]\right)$.
This is also the necessary requirement for the definition of the
plus-function 
\ba{plus}
\braket{(1-x)^{-1}_+}{f}=\int\limits_0^1 dx\;\frac{f(x)-f(1)}{1-x}
\ea 
which will be of importance later.  
The singular behaviour of the splitting function at $x\uparrow1$
corresponds to high moments.
But for large $n$ the weight functions $x^n$ form a Cauchy series with
respect to the norm of the Schwartz-Sobolev space 
${\mathfrak S}_{1}\left([0,1]\right)$
\be{schaetz}
||x^{n+1}-x^n||^2_{{\mathfrak S}_{1}} = {\cal O}\left(\frac{1}{n}\right)
\,.
\ee
The convergence is even stronger (${\cal O}(1/n^3)$) for the
$L_2$-norm, but, e.g., it does not persist for the 
${\mathfrak S}_{2}$-norm.
This Schwartz-Sobolev space ${\mathfrak S}_{2}\left([0,1]\right)$
is the domain of definition of the derivative of the
$\delta$-distribution $\braket{\delta_\chi'}{f}=-f'(\chi)$. 
If we now assume that the singularity structure of the splitting
function does not contain such distributions, but only the
Dirac-$\delta$ or the plus-function, we may restrict ourselves to the
(lowest) Schwartz-Sobolev space ${\mathfrak S}_{1}\left([0,1]\right)$.  
Then the inequality of Schwarz 
\ba{Ungl}
\braket{g}{f} \leq 
||g||_{{\mathfrak S}_{-1}}
||f||_{{\mathfrak S}_{1}}
\ea
implies that
\be{grossemoments}
\int\limits_0^1 dx\, x^{n+1} {\rm Abs}\,T -
\int\limits_0^1 dx\, x^n {\rm Abs}\,T
\rightarrow
0
\ee
holds for $n\uparrow\infty$.
Note that the contribution of $\delta(1-x)$ to the moments is a constant.
From Eq.~\gl{grossemoments} the behavior of the anomalous
dimension for $n\uparrow\infty$ can be determined by the even
moments only.
The singular part $T_{\rm sing}$ of $T$ corresponds to high moments,
so that it is already fixed in this way
(as long as it can be defined within ${\mathfrak S}_{-1}$).

Now let us analyze the regular part $T_{\rm reg}$ of the forward
scattering amplitude $T=T_{\rm reg}+T_{\rm sing}$. 
Since $T$ is assumed to be an analytic function, we suppose that its
moments possess an analytic continuation into the complex plane,
i.e. we presume that
\ba{analyt}
M(s)=\int\limits_0^1 dx \, x^s {\rm Abs}\,T_{\rm reg}
\ea
is an analytic function for $s>1$. 
This assumption is very reasonable but not necessary for our
main result, which mainly concerns the singular part. 
After a conformal mapping of the complex plane onto the unit sphere
${\mathbb S}_2$ the known even moments describe a convergent series   
of the (supposed to be) analytic function $M(s)$.
Since we have subtracted the singular part -- which is completely
determined by the behaviour of the high moments -- the remaining
regular part approaches zero in this limit 
\be{regulaer}
\int\limits_0^1 dx\,x^n {\rm Abs}\,T_{\rm reg} \rightarrow 0
\,.
\ee
Now the theorem of identity for analytic functions can be applied.
(If two analytic functions coincide at a convergent series, then they
coincide everywhere.) 
It follows that the result for even moments remains correct for odd
moments as well. This means that we get all moments by calculating the
even moments only (under the assumptions made).

We require the same assumptions to hold for $h_1$ and proceed in the
same way.

\subsection{Results for $g_1$}

The forward Compton scattering amplitude corresponding to the polarized
structure function $g_1$ reads after expansion in $\omega=1/x$:
\be{g1omega}
\left.\phantom{\frac12}
T_{\mu\nu} 
\right|_{\mbox{$\displaystyle
\frac{i\varepsilon_{\mu\nu\lambda\rho} q^\lambda s^\rho}
{p\cdot q}$}}
\;=\;
2i\varepsilon_{\mu\nu\lambda\rho} \frac{q^\lambda s^\rho}{p \cdot q}
\sum_{n=0}^\infty
C_{g_1,n} A_{g_1,n}
\,\omega^{n+1}
\,.
\ee
We have computed all one-gluon-exchange diagrams mentioned above
with the vertex $\Gamma_\nu=\gamma_\nu$. In the framework of
dimensional
regularization with $d=4-\eps$ the divergent part appears
as coefficient of $2/\eps$.
In order to analyze the calculation we present the results for
all graphs separately. After projection onto the appropriate
Lorentz structure
\be{kurz}
\alpha_{\mu\nu}\;=\;
\frac{2i\varepsilon_{\mu\nu\lambda\rho}q^\lambda s^\rho}{p\cdot q}
\ee
we get using the results from the Appendix
\ba{g1self}
{\cal M}_{S,\mu\nu}&=&-C_F \alpha_{\mu\nu}
\frac2\eps\sum\limits_{n=0}^\infty\omega^{n+1}
\,.
\ea
Note that this geometrical series can be summed up to
yield $\omega/(1-\omega)$ in the unphysical regime $\omega<1$.
For the physically interesting absorptive part
a $\delta(1-x)$-distribution is generated. 
The results for the vertex
corrections and the boxgraph read
\ba{g1vertexr}
{\cal
M}_{V1,\mu\nu}&=&C_F \alpha_{\mu\nu}
\frac2\eps\sum\limits_{n=0}^\infty\omega^{n+1}
\left(1+\sum\limits_{k=1}^n\frac2{k+1}\right)
\,,
\ea
\ba{g1vertexl}
{\cal
M}_{V2,\mu\nu}&=&C_F \alpha_{\mu\nu}
\frac2\eps\sum\limits_{n=0}^\infty\omega^{n+1}
\left(1+\sum\limits_{k=1}^n\frac2{k+1}\right)
\,,
\ea
\ba{g1box}
{\cal
M}_{B,\mu\nu}&=&-C_F \alpha_{\mu\nu}
\frac2\eps\sum\limits_{n=0}^\infty\omega^{n+1}
\frac{2}{(1+n)(2+n)}
\,.
\ea
The gluon propagator was taken in Feynman-gauge
so that it is
reduced to the $g_{\mu\nu}$-term. 
Nevertheless, we checked that
the sum of the contributions 
arising from the $k_\mu k_\nu$-term
vanishes as required by gauge invariance.

Taking these results together we find
\ba{g1sum}
{\cal M}_{\mu\nu} 
&=& C_F\,\alpha_{\mu\nu}\frac2\eps
\sum\limits_{n=0}^\infty\omega^{n+1} \times \nn
&&
\left(\frac4{1+n}-\frac2{(1+n)(2+n)}+4 S_n-3\right)
\,,
\ea
where
\be{sn}
S_n = \sum\limits_{k=1}^n\frac1k
\,.
\ee
The leading order anomalous dimension becomes
\be{g1anom}
\gamma_{g_1,n}
=\frac{C_F}{(4\pi)^2}
\left(\frac4{1+n}-\frac2{(1+n)(2+n)}+4 S_n-3\right)
\,,
\ee
and the inverse Mellin transformation generates
\be{g1splitting}
P_{g_1}(x)=\frac{C_F}{(4\pi)^2}
\left(4-2(1-x)-\frac4{(1-x)_+} -3\delta(1-x)\right)
\ee
for the splitting function. This result is well
known in literature \cite{Cra83}. Usually, the distribution
$3\delta(1-x)$ is fixed by the sum rule
\be{g1sumrule}
\int\limits_0^1 dx\,P_{g_1}(x) \;=\; 0
\ee
and is not calculated directly (as we did here by
using the dispersion relation Eq.~\gl{disp}). 
Due to conservation of helicity
the integral over the
quark distribution does not change with $Q^2$, 
i.e.~$\gamma_{g_1,0}=0$.
As transversity is not conserved 
a corresponding sum rule for the transversity
distribution is not known.

\subsection{Results for $h_1$}

In the same way as for the polarized structure function
we expand Eq.~\gl{h1Tmu} in $\omega$:
\be{h1omega}
\left.
T_\mu 
\right|_{\mbox{$\displaystyle s_{\bot\mu}$}}
\;=\;
s_{\perp\mu}
\sum_{n=0}^\infty
C_{\perp,n} A_{\perp,n} \omega^{n+1}
\,.
\ee
In this case we find for the one gluon exchange graphs 
(again using the results from the Appendix) after projection
on $s_{\perp\mu}$
\ba{h1self}
{\cal M}_{S,\mu}&=&-C_F\,s_{\perp\mu}\frac2\eps\sum\limits_{n=0}^\infty\omega^{n+1}
\,,
\ea
\ba{h1vertexr}
{\cal
M}_{V1,\mu}&=&C_F\,s_{\perp\mu}\frac2\eps\sum\limits_{n=0}^\infty\omega^{n+1}
\left(4+\sum\limits_{k=1}^n\frac2{k+1}\right)
\,,
\ea
\ba{h1vertexl}
{\cal
M}_{V2,\mu}&=&C_F\,s_{\perp\mu}\frac2\eps\sum\limits_{n=0}^\infty\omega^{n+1}
\left(1+\sum\limits_{k=1}^n\frac2{k+1}\right)
\,,
\ea
\ba{h1box}
{\cal
M}_{B,\mu}&=& 0
\,.
\ea
The boxgraph does not lead to any divergent contribution
to the Lorentz structure $s_{\perp\mu}$.

Adding up all graphs we get
\be{h1sum}
{\cal M}_{\mu}= C_F\,s_{\perp\mu}\frac2\eps\sum\limits_{n=0}^\infty\omega^{n+1}
\left(\frac4{1+n}+4S_n\right)
\,.
\ee
This leads to the anomalous dimension
\be{h1anom}
\gamma_{\perp,n}
=\frac{C_F}{(4\pi)^2}\left(\frac4{1+n}+4 S_n\right)
\ee
and after an inverse Mellin transformation to
the splitting function:
\be{h1splitting}
P_\perp(x)=\frac{C_F}{(4\pi)^2}\left(4-\frac4{(1-x)_+}\right)
\,.
\ee
This is the splitting function corresponding to the operator defined
in Eq.~\gl{h1Tmu}. Additionally, according to the method of
Ioffe and Khodjamirian, this result may be identified with the
splitting function of the (twist-2) transversity distribution as defined
in Eq.~\gl{defoperh1} in the Appendix. 
However, our result Eq.~\gl{h1splitting}
differs from the one found in literature for the transversity
distribution \cite{Art90} 
\be{h1lit}
P_\perp^{\rm lit}(x)=\frac{C_F}{(4\pi)^2}\left(4-\frac4{(1-x)_+}
-3\delta(1-x)\right)
\ee
at $x=1$.

\section{Discussion}

The splitting functions Eqs.~\gl{g1splitting} 
and \gl{h1splitting} for $g_1$ and $h_1$, respectively,
were calculated using the same method. In the
case of the operator corresponding to the longitudinally polarized
structure function the expected result was found. Nevertheless,
the analogous calculation for the transversity operator leads
to a result which differs from the one in literature.
We want to elucidate possible reasons for this discrepancy.

\subsection{Sum rules}

As emphasized in Sec.~\ref{secdisp} a direct calculation of the
splitting function without using the dispersion relation
Eq.~\gl{disp} does not fix the behavior at $x=1$. This is done
by the sum rule Eq.~\gl{g1sumrule} in the case of the helicity asymmetry
operator.
We showed above, that the dispersion relation leads to exactly
the result expected from the sum rule without any further
argument.

One may observe that the splitting functions
for the helicity asymmetry and the transversity operator
Eqs.~\gl{g1splitting} and \gl{h1splitting}
differ by an additional term proportional to $1-x$
which vanishes for $x=1$ and an additional
$\delta(1-x)$. This distribution enters $P_{g_1}(x)$
with the same weight as the missing term
found in literature Eq.~\gl{h1lit} for $P_\perp$.
This is not accidental. 
As for the transversity 
operator no sum rule corresponding to Eq.~\gl{g1sumrule}
is known, the behavior for $x=1$ of the
transversity splitting function
was fixed with the general argument 
\cite{Art90,Bar97,Hay97}, that the
probability of a transversity flip should vanish
for $x=1$. It follows from this (reasonable) requirement
that the splitting functions $P_\perp$ and $P_{g_1}$
should behave identically at $x=1$. We want to emphasize
that by using this argument, the divergent structure
at $x=1$ is not calculated explicitly but fixed by hand.

\subsection{Comparing both calculations}

The Feynman graphs contributing to the splitting functions
differ solely by the vertex $\Gamma_\nu$ which is $\gamma_\nu$
in the case of the helicity asymmetry operator and
unity for the transversity operator. For this reason the divergent
part of one gluon exchange contributions are equal for 
the self energy ${\cal M}_S$
as well as for the vertex correction ${\cal M}_{V2}$ which do
not involve the differing vertex into the loop integral.
Comparing the explicitly calculated
results Eqs.~\gl{g1self} and \gl{g1vertexl}
with Eqs.~\gl{h1self} and \gl{h1vertexl}, respectively,
this statement is confirmed.

Looking at the results for the boxgraph Eqs.~\gl{g1box} and \gl{h1box}
we observe a difference. The boxgraph is not divergent for the
transversity operator due to an extra factor $4-d$
appearing by projection on the corresponding Lorentz structure.
After an inverse Mellin transformation the contribution
Eq.~\gl{g1box} to the anomalous dimension is translated into
the $1-x$ term in Eq.~\gl{g1splitting}. This term vanishes
for $x=1$ and for this reason is not a good candidate to
explain the difference at $x=1$.

In this way only the vertex correction ${\cal M}_{V1}$
with a gluon exchanged at the $\gamma_\nu$ or the unity
vertex, respectively, may be responsible for the difference
at $x=1$. Comparing Eqs.~\gl{g1vertexr} and \gl{h1vertexr}
the result differ by an additional constant contribution
to the anomalous dimension. An inverse Mellin transformation
of a constant leads exactly to the additional term
$-3\delta(1-x)$ for the polarized splitting function
$P_{g_1}$. It is plausible that two graphs with one gluon
exchanged at different vertices do not lead to the same result.
Anyhow, we will have to look for a conceptual problem if
we want to save the general statement of vanishing transversity
flip probability at $x=1$.

\subsection{Assumptions}

We based the identification of the calculated anomalous
dimension with the one of the transversity distribution
on the argument of Ioffe and Khodjamirian \cite{Iof95}.
Here it was shown, that the operator used in the forward
scattering amplitude Eq.~\gl{h1Tmu} indeed has a relation
to the definition of the transversity distribution Eq.~\gl{defoperh1}.
Their argument remains correct on the leading twist-level, which
is appropriate to our purpose.
The identification may be affected by anomalies. But these
appear first in second order in the strong coupling. This
implies that our first order calculation remains unaffected
from possible anomalies.

In addition the method of Ioffe and Khodjamirian is based on 
light cone dominance. One may contest, that this assumption is 
not appropriate to describe the splitting
function at $x=1$. Here a potentially unbounded number of 
zero-momentum gluons appears, which are not well described 
by free quark operators. 
Nevertheless, the
identification of the structure function (defined in 
Eq.~\gl{Teqh1}) with the transversity distribution
(defined in Eq.~\gl{defoperh1}) on the leading twist level
and in leading order of perturbation theory is unaffected by
the above consideration, so that the
leading order anomalous dimension should be reproduced correctly.
Consequently, 
if the method of Ioffe and Khodjamirian is applicable to
the problem under consideration, we will have to look
for other reasons for the discrepancy.

Our calculation of the anomalous dimension for $g_1$ and $h_1$
rely both on the assumption that ${\rm Abs}\,T$ in Eq.~\gl{disp}
is element of the Schwartz-Sobolev space ${\mathfrak S}_{-1}$. 
The result Eq.~\gl{g1anom} for $g_1$ is in accordance with literature
which is not the case for $h_1$, so that
the above assumption may be questioned for the transversity operator.
If ${\rm Abs}\,T_\mu$ is not element of the Schwartz-Sobolev space
${\mathfrak S}_{-1}$, the irregular part contains other
contributions than a $\delta$-distribution. 
For example derivatives of $\delta$ could appear. 
As the difference of the results Eqs.~\gl{h1splitting} and \gl{h1lit} 
is element of ${\mathfrak S}_{-1}$, it cannot be explained in this way.
This means that the reason for the discrepancy
is not related to our assumptions.

The assumption that the regular part of the forward scattering
amplitude is an analytic function, is completely independent
of the problem at $x=1$. Analyticity is necessary to
ensure that our calculation of the terms with $x<1$ 
reproduces the correct regular part of the splitting function.

\subsection{Implications}

It follows from the analysis of the assumptions that our result
for the transversity splitting function can be interpreted in
two alternative ways: First,
the correspondance of Eqs.~\gl{Teqh1} and \gl{defoperh1}
shown by Ioffe and Khodjamirian
contains a conceptual difficulty at $x=1$.
Alternatively, our result implies a non-vanishing
probability of transverity flip by emitting infrared gluons
from quarks.
This is really hard to believe. Especially, the number 
of infrared gluon emissions is arbitrary large so that
the transversity becomes completely statistically
distributed. On the other hand the regularization of
the infrared regime was done by hand, exactly in order
to avoid this surprising interpretation. It was not
deduced from first principles of QCD. 
Our calculation shows that (under the assumptions
we made) the vanishing transversity flip probability
for zero-momentum gluons does not follow analytically.

\section*{Appendix}

\subsection*{A.~Distribution functions}

The polarized quark distribution is defined in leading twist by
\be{defoperg1}
\int \frac{d\lambda}{4\pi}\,e^{i\lambda x}
\bra{p,s} \overline{\psi}(0) \gamma_{\mu} \gamma_5 \psi(\lambda n)
\ket{p,s}
=
g_1(x,q^2) p_\mu s\cdot n
\,.
\ee
Again $x=-q^2/(2p\cdot q)$ is the Bjorken variable.
$n^\nu$ is a light cone vector with $n^2=0$ of dimension
$(mass)^{-1}$, $p$ is the proton four-momentum with $p^2=m^2$
and $p\cdot n=1$,
and $s$ is the proton spin four-vector with $s^2=-1$.
In deep inelastic electron proton scattering the distribution $g_1$ 
describes the asymmetry of nucleons polarized parallel or antiparallel
to the longitudinally polarized lepton.

The definition of $h_1$ in terms of an operator matrix element reads
\cite{Ral79,Jaf91}:
\ba{defoperh1}
&&\int \frac{d\lambda}{4\pi}\,e^{i\lambda x}
\bra{p,s} \overline{\psi}(0) \sigma_{\mu\nu} i\gamma_5 \psi(\lambda n)
\ket{p,s}\nn
&&=\,
h_1(x,q^2)\left(s_{\bot\mu}p_\nu - s_{\bot\nu}p_\mu\right)
\,,
\ea
where the transverse
part of the spin vector is defined by the decomposition
$s_\mu = (s\cdot n)p_\mu + (s\cdot p)n_\mu + s_{\bot\mu}$.

\subsection*{B.~Integrals}

The loop integrals neccessary for the calculation of the Feynman
diagrams in Sec.\,\ref{SplitCalc} possess infrared as well as
ultraviolet divergences, which have to be regularized. For our
purposes we used the dimensional regularization scheme with
$d=4-\eps$. Since we are interested in the anomalous dimension and
therefore in the divergent parts only, we can evaluate the integrals
in the $\eps\to 0$ limit. For the integral appearing in the
selfenergy graph we find
\begin{eqnarray}
I_S^\alpha &=& i (4\pi)^2 \mu^{4-d} \times\nn
&&\int
\frac{d^dk}{(2\pi)^d}\frac{(p+q-k)^\alpha-i\eta}{\left[(p+q-k)^2+\eta^2\right] 
(k^2+i\eta)}\nn
&=& - \frac2\eps[(p+q)^\alpha-i\eta](1+{\cal O}(\eps))
\,.
\end{eqnarray}
Both vertex corrections involve the same integral
\begin{eqnarray}
\nonumber
I_V^{\alpha\beta} &=& i (4\pi)^2 \mu^{4-d}\int \frac{d^dk}{(2\pi)^d}
\times \\
&&
\frac{[(p-k)^\alpha-i\eta][(p+q-k)^\beta-i\eta]}{[(p+q-k)^2+\eta^2]
[(p-k)^2+\eta^2](k^2+i\eta)}\nn
&=& - \frac2\eps \frac{g^{\alpha\beta}}4(1+{\cal O}(\eps))
\,.
\end{eqnarray}
The integral corresponding to the boxgraph
\begin{eqnarray}
&&I_B^{\alpha\beta\lambda} = i (4\pi)^2 \mu^{4-d}\int\frac{d^dk}{(2\pi)^d}
\times
\nonumber \\
&&
\frac{[(p-k)^\alpha-i\eta][(p+q-k)^\beta-i\eta][(p-k)^\lambda-i\eta]}
     {[(p+q-k)^2+\eta^2][(p-k)^2+\eta^2]^2 (k^2+i\eta)}
\nonumber \\
\end{eqnarray}
is obviously infrared divergent only, which leads
to a maximum divergence of $1/\eps$. In the case of the transversal structure
function $h_1$ the boxgraph is supplied with a prefactor of $4-d$
resulting from the according Lorentz structure. Therefore it yields no
contributions to the anomalous dimension. On the other hand the
boxgraph leads to a term $\sim(1-x)$ (see Eq.\,\gl{g1box}) for the polarized 
structure
function $g_1$. However, since we are mainly interested in the behavior of
the splitting functions in the limit $x\uparrow 1$, 
the boxgraph is not relevant 
for our purposes.  

\subsection*{Acknowledgment} 

We thank X. Artru, B.L. Ioffe, G. Soff, and L. Szymanowski   
for enlightening discussions.
This work was partially supported by BMBF, DFG, and GSI.

% ===================================================================

\end{document}